\documentstyle[12pt]{article}

\begin{document}
\pagestyle{empty}

\begin{center}
{\bf A PATH INTEGRAL APPROACH TO CURRENT }

\vspace{1cm} A. Marchewka\\[5mm]
Department of Physics\\[0pt]
Tel-Aviv University\\[0pt]
Ramat-Aviv, Tel-Aviv 69978, Israel\\
e-mail marhavka@ccsg.tau.ac.il\\[5mm]
Z. Schuss\\[5mm]
Department of Mathematics\\[0pt]
Tel-Aviv University\\[0pt]
Ramat-Aviv, Tel-Aviv 69978, Israel\\
e-mail schuss@math.tau.ac.il\\\vspace{1cm}
{\bf{TAUP 2570-99}}\\
\vspace{1cm} {\bf ABSTRACT}
\end{center}

\vspace{5mm} 
Discontinuous initial wave functions or wave functions with
discontintuous derivative and with bounded support arise in a
natural way in various situations in physics, in particular in
measurement theory. The propagation of such initial wave
functions is not well described by the
Schr\"odinger current which vanishes on the
boundary of the support of the wave function. This propagation
gives rise to a uni-directional current at the boundary of the
support. We use path integrals to define current and
uni-directional current and give a direct derivation of the
expression for current from the path integral formulation for
both diffusion and quantum mechanics. Furthermore, we give an
explicit asymptotic expression for the short time propagation
of  initial wave function with compact support for both the
cases of discontinuous derivative and discontinuous wave
function. We show that in the former case the probability
propagated across the boundary of the support in time $\Delta
t$ is $O\left( \Delta t^{3/2}\right) $ and the initial
uni-directional current is $O\left( \Delta t^{1/2}\right) $.
This recovers the Zeno effect for continuous detection of a
particle in a given domain. For the latter case the
probability propagated across the boundary of the support in
time $\Delta t$ is $O\left( \Delta t^{1/2}\right) $ and the
initial uni-directional current is $O\left( \Delta
t^{-1/2}\right) $. This is an anti-Zeno effect. However, the
probability propagated across a point located at a finite
distance from the boundary of the support is $O\left( \Delta
t\right) $. This gives a decay law.
\newpage

\pagestyle{plain} \setcounter{page}{1}

\noindent
{\bf 1. Introduction}\newline
\renewcommand{\theequation}{1.\arabic{equation}} \setcounter{equation}{0}

Detection of the coordinate of a quantum particle can give the
negative result that the particle is not in a given domain, for
example, it is not in a half space bounded by a given plane. This means
that at the moment of measurement the wave function of the undetected
particle vanishes identically in the given domain, for example, beyond the
separating plane. It follows that the wave function of the particle is supported
outside the measured domain. The boundary of the support is the
boundary of the domain, for example, it can be the plane separating the
measured half space from the other half. 
The instantaneous killing of the wave function in a domain $G$ for a
given time interval $[t_0,t_0+\Delta t]$ can be accomplished mathematically by 
introducing in the Schr\"odinger equation the (dimensionless) time 
dependent potential
\begin{eqnarray*}
V(x,t)=\infty\cdot\chi_G(x)\chi_{[t_0,t_0+\Delta t]}(t)
\end{eqnarray*}
where 
\begin{eqnarray*}
\chi_G(x)=\left\{\begin{array}{ll}1&\mbox{if $x\in
G$}\\0&\mbox{otherwise.}\end{array}\right.
\end{eqnarray*}
After the instantaneous negative detection the particle in the time
interval $[t_0,t_0+\Delta t]$ it
is free to propagate into the domain, that is, the potential $V(x,t)$ is
turned off instantaneously. A discontinuous wave function with compact
support can be realized mathematically by introducing a similar
potential that contains the spatial derivative of the delta function on
te boundary of the domain and an infinite potential inside the domain.

The propagation of the wave function of the particle after the 
negative measurement is not well described by the Schr\"odinger current 
because the compactly supported initial wave function after the 
measurement gives rise to an initially vanishing Schr\"odinger current
at the boundary of the support. Indeed, assuming the wave function is
continuous, since
\begin{eqnarray*}
\left.\psi(x,t)\right|_{\partial G}=0,\quad\mbox{for all $t_0\leq t\leq
t_0+\Delta t$},
\end{eqnarray*}
where $t_0$ is the instant of the measurement and $\partial G$ is the
boundary of the measured domain, we must have
\begin{eqnarray*}
J_{\partial G}(x,t)=\frac \hbar m\Im \mbox{m\thinspace }\psi
(x,t)\nabla\bar{\psi}(x,t)=0,\quad\mbox{for all $t_0\leq t\leq 
t_0+\Delta t$},
\end{eqnarray*}
It is obvious, however, that despite the vanishing current on $\partial
G$ there is propagation across $\partial G$ after the instant $t_0+\Delta
t$. This is in general the case in Schr\"odinger's equation if the initial
wave function has compact support. The case of a discontinuous initial
wave function is problematic as well.

It is the purpose of this paper to calculate the short time propagation
of the wave function across the boundary of the support of the initial
wave function. This propagation gives rise to an instantaneous
uni-directional current into the measured domain $G$. Our main result
is an explicit asymptotic expression for the wave function at time
$t_0+\Delta t$ for small $\Delta t$.  Our
analysis is one-dimensional, however the generalization to higher
dimensions is straightforward. 

>From this expression, we find that
for a continuous wave function the probability that propagates 
across $\partial G$ in time $\Delta t$ is
$O\left(\Delta t^{3/2}\right)$ and the rate of propagation is
$O\left(\Delta t^{1/2}\right)$. The latter means that the initial
uni-directional current is $O\left(\Delta t^{1/2}\right)$. 
We also determine the probability $P_c$
that propagates in time $\Delta t$ beyond a distance $c$ from $\partial G$,
\begin{eqnarray*}
P_c=O\left(\left(\frac{\Delta t}{c}\right)^3\right).  
\end{eqnarray*}
This expansion is valid for $c\geq\sqrt{\alpha}$. For example, if 
$c=O\left(\Delta t^{1/3}\right)$, we obtain $P_c=O\left(\Delta t^2\right)$. For 
$c=O\left(\sqrt{\Delta t}\right)$ the propagated probability is $O\left(\Delta
t^{3/2}\right)$. This gives an estimate on the dependence of the probability
of detection on the size of the detector. Detection at a point or in a
domain is discussed in the context of absorption in separate papers
\cite{PLA}-\cite{measurement}.

Our result recovers the Zeno effect \cite{Peres,Grigolini} that a quantum
particle cannot arrive at a point under continuous observation (the
particle is ``frozen'' or ``reflected back'' by the continuous
observation). This result is in agreement with the result of \cite{Aharonov}
(see Section 6 for more detailed discussion).

For a discontinuous initial wave function, we find that 
the probability that propagates across $\partial G$ in time $\Delta t$ is
$O\left(\Delta t^{1/2}\right)$ and the rate of propagation is
$O\left(\Delta t^{-1/2}\right)$. The latter means that the initial
uni-directional current is $O\left(\Delta t^{-1/2}\right)$ and is initially
infinite. This result is an anti-Zeno effect \cite{Kaulakys}. The probability
$P_c$ for this case is given by
\begin{eqnarray*}
P_c=O\left(\frac{\Delta t}{c}\right).
\end{eqnarray*}
This implies that the survival probability decays exponentially 
(see eq.(\ref{S(t)})).

Wave functions with compact support and discontinuous derivatives appear
in other applications as well \cite{PHD,Cheon}. They represent actual physical
situations and are of both theoretical and practical interest. 

The notion of uni-directional current can be used to define the notion
of ``time of arrival'' and measurement in quantum mechanics. A review of
different approaches to the definition of uni-directional current related to
time of arrival in quantum mechanics is given in \cite{Delgado1,Muga3}. 
Our approach to the definition of a uni-directional current is different 
from all the above mentioned approaches. It is defined only for wave
functions that vanish identically beyond the detector, that is, 
its support is initially bounded by the detector and is defined only at the
boundary of the support. At such points the uni-directional current
and the current are the same.

Out approach to the calculation of the  uni-directional current 
is based on a Feynman integral approach. This seems to be a new
approach that is fundamentally different than the above mentioned approaches
in that it is not approximate and gives the full asymptotic behavior of the
wave function for short times. A direct derivation of the Schr\"odinger
current from the Feynman integral is given and extended to the case of a
uni-directional current.

In this approach uni-directional currents appear when the wave function
satisfies the following conditions,

\begin{itemize}
\item  The initial wave function has compact support.

\item  At the boundary of the support either the wave function or its
derivative has a discontinuity.
\end{itemize}

Initial wave functions that satisfy these conditions are denoted ICSWF
(initial compact support wave function). 
In general, the solutions of Schr\"{o}dinger's equation do not develop
discontinuities or non-smoothness. However, there are situations where
non-smoothness does occur \cite{Cheon}. If an infinite potential
barrier is introduced, the gradient of the wave function has a discontinuity
across the barrier. Indeed, in our formalism for the description of an
absorbing wall such a discontinuity arises in a natural way.

In Sections 2, the case of compact support in diffusion theory is
reviewed. The Wiener path integral formulation of diffusion is shown to
lead to the classical definitions of diffusion current and to a
uni-directional diffusion current at an absorbing boundary. This
derivation seems to be new. In Section 3, a Feynman integral
formulation is used to define quantum probability current and a
new direct derivation of the Schr\"odinger current from Feynman's integral is
given. This definition leads to the definition of a uni-directional
current at the boundary of the support of the initial wave function. In
Section 4, the main results of the paper are presented. These are short
time asymptotic expansions for the propagation of an ICSWF and the
uni-directional current. Finally, Section 6 contains a discussion and
summary of the results.  \newline

\noindent
{\large {\bf 2. Current and uni-directional current in diffusion}}\newline
\renewcommand{\theequation}{2.\arabic{equation}} \setcounter{equation}{0}

We consider a diffusion process, $x(t)$, with noise coefficient $\sigma
(x,t) $ and drift $b(x,t)$ \cite{book}. The transition probability density
of the process is denoted $p(y,t\,|\,x,s)$. This is the probability density 
\[
p(y,t\,|\,x,s)=\frac \partial {\partial y}\Pr \left\{
x(t)<y\,|\,x(s)=x\right\} . 
\]
It satisfies the Fokker-Planck equation 
\begin{eqnarray}
\frac{\partial}{\partial t}p(y,t\,|\,x,s)= \frac{1}{2}\frac{\partial^2}{%
\partial y^2}\left[\sigma^2(y,t) p(y,t\,|\,x,s)\right]-\frac{\partial}{%
\partial y}\left[b(y,t) p(y,t\,|\,x,s)\right]  \label{FPEQ}
\end{eqnarray}
with the initial condition 
\begin{eqnarray}
\lim_{t\downarrow s}p(y,t\,|\,x,s)=\delta (y-x).  \label{IC1}
\end{eqnarray}
At an absorbing boundary at $y=0$, say, the function $p(y,t\,|\,x,s)$
satisfies the boundary condition 
\begin{eqnarray}
p(0,t\,|\,x,s)=0.  \label{abc}
\end{eqnarray}

The uni-directional probability current (flux) density at a point $x_1$ is
the probability per unit time of diffusing trajectories that propagate from
the ray $x<x_1$ into the ray $x>x_1$. It is therefore given by 
\begin{equation}
J_{LR}(x_1,t)=\lim_{\Delta t\rightarrow 0}J_{LR}(x_1,t,\Delta t),
\label{JLRN}
\end{equation}
where 
\begin{eqnarray}
&&J_{LR}(x_1,t,\Delta t)=  \label{JLRDt} \\
&&\frac 1{\Delta t}\int_{x_1}^\infty dx\,\int_{-\infty }^{x_1}\frac{dy}{%
\sqrt{2\pi \Delta t}\,\sigma (y,t)}\exp \left\{ -\frac{\left[
x-y-b(y,t)\Delta t\right] ^2}{2\sigma ^2(y,t)\Delta t}\right\} p(y,t-\Delta
t)  \nonumber
\end{eqnarray}
(note that the dependence of $p$ on the backward variables $x,s$ has been
suppressed). The integral (\ref{JLRDt}) can be calculated by the Laplace
method \cite{Bender} at the saddle point $x=y=x_1$. First, we change
variables in (\ref{JLRDt}) to 
\[
x=x_1+\xi \sqrt{\Delta t},\quad y=x_1-\eta \sqrt{\Delta t} 
\]
to obtain 
\begin{eqnarray*}
&&J_{LR}(x_1,t,\Delta t)=\int_0^\infty d\xi \,\int_0^\infty \frac{d\eta }{%
\sqrt{2\pi \Delta t}\,\sigma (x_1-\eta \sqrt{\Delta t},t)}\times \\
&&\exp \left\{ -\frac{\left[ \xi +\eta -b(x_1-\eta \sqrt{\Delta t},t)\sqrt{%
\Delta t}\right] ^2}{2\sigma ^2(x_1-\eta \sqrt{\Delta t},t)}\right\} p\left(
x_1-\eta \sqrt{\Delta t},t-\Delta t\,\right) \,
\end{eqnarray*}
and changing the variable in the inner integral to $\eta =\zeta -\xi $, we
get 
\begin{eqnarray}
&&J_{LR}(x_1,t,\Delta t)=\int_0^\infty d\xi \,\int_\xi ^\infty \frac{d\zeta 
}{\sqrt{2\pi \Delta t}\,\sigma (x_1-\left( \zeta -\xi \right) \sqrt{\Delta t}%
,t)}\times  \label{jxieta} \\
&&\exp \left\{ -\frac{\left[ \zeta -b(x_1-\left( \zeta -\xi \right) \sqrt{%
\Delta t},t)\sqrt{\Delta t}\right] ^2}{2\sigma ^2(x_1-\left( \zeta -\xi
\right) \sqrt{\Delta t},t)}\right\} p\left( x_1-\left( \zeta -\xi \right) 
\sqrt{\Delta t},t-\Delta t\,\right) .  \nonumber
\end{eqnarray}
Next, we expand the exponent in powers of $\sqrt{\Delta t}$ to obtain 
\begin{eqnarray}
&&\frac{\left[ \zeta -b(x_1-\left( \zeta -\xi \right) \sqrt{\Delta t},t)%
\sqrt{\Delta t}\right] ^2}{2\sigma ^2(x_1-\left( \zeta -\xi \right) \sqrt{%
\Delta t},t)}=  \label{exponent} \\
&&\frac{\zeta ^2}{2\sigma ^2(x_1,t)}+\left[ \frac{\zeta ^2\left( \zeta -\xi
\right) \sigma ^{2\prime }(x_1,t)}{2\sigma ^4(x_1,t)}-\frac{\zeta b(x_1,t)}{%
\sigma ^2(x_1,t)}\right] \sqrt{\Delta t}+O\left( \Delta t\right) ,  \nonumber
\end{eqnarray}
the pre-exponential factor, 
\begin{equation}
\frac 1{\,\sigma \left( x_1-\left( \zeta -\xi \right) \sqrt{\Delta t}%
,t\right) }=\frac 1{\,\sigma (x_1,t)}\left[ 1+\frac{\sigma ^{\prime }(x_1,t)%
}{\sigma (x_1,t)}\left( \zeta -\xi \right) \sqrt{\Delta t}+O\left( \Delta
t\right) \right] ,  \label{prexp}
\end{equation}
and the pdf 
\begin{eqnarray}
&&p\left( x_1-\left( \zeta -\xi \right) \sqrt{\Delta t},t-\Delta t\,\right) =
\label{pNexp} \\
&&p(x_1,t)-\frac{\partial p(x_1,t)}{\partial x}\left( \zeta -\xi \right) 
\sqrt{\Delta t}-\frac{\partial p(x_1,t)}{\partial t}\xi \Delta t+O\left(
\Delta t^{3/2}\right) .  \nonumber
\end{eqnarray}
Using the expansions (\ref{exponent})-(\ref{pNexp}) in (\ref{jxieta}), we
obtain 
\begin{eqnarray}
&&J_{LR}(x_1,t,\Delta t)=  \label{JLRexp} \\
&&\int_0^\infty d\xi \int_\xi ^\infty \frac{d\zeta }{\sqrt{2\pi \Delta t}%
\,\sigma (x_1,t)}\exp \left\{ -\frac{\zeta ^2}{2\sigma ^2(x_1,t)}\right\}
\,\left\{ p\left( x_1,t\right) -\frac{\partial p\left( x_1,t\right) }{%
\partial t}\Delta t\right. {}{}{}  \nonumber \\
&&-\sqrt{\Delta t}\left[ \left( \frac{\zeta ^2\left( \zeta -\xi \right)
\sigma ^{\prime }\left( x_1,t\right) }{\sigma ^3\left( x_1,t\right) }-\frac{%
\zeta b(x_1,t)}{\sigma ^2\left( x_1,t\right) }-\frac{\sigma ^{\prime }\left(
x_1,t\right) \left( \zeta -\xi \right) }{\sigma \left( x_1,t\right) }\right)
p\left( x_1,t\right) \right.  \nonumber \\
&&\left. \left. \frac {}{}+\left( \zeta -\xi \right) p^{\prime }\left(
x_1,t\right) \right] +O\left( \Delta t\right) \right\} .  \nonumber
\end{eqnarray}
Similarly, 
\[ J_{RL}(x_{1},t)=\lim_{\Delta t\rightarrow 0}J_{RL}(x_{1},t,\Delta t),\]
where 
\begin{eqnarray}
&&J_{RL}(x_1,t,\Delta t)=  \label{JRLDt} \\
&&\frac{1}{\Delta t}\int_{-\infty}^{x_1}dx\,\int_{x_1}^\infty 
\frac{dy}
{\sqrt{2\pi \Delta t}\,\sigma (y,t)}\exp \left\{ -\frac{\left[
x-y-b(y,t)\Delta t\right] ^{2}}{2\sigma ^{2}(y,t)\Delta t}\right\}
p(y,t-\Delta t).\nonumber
\end{eqnarray}
The change of variables in (\ref{JRLDt})
\[ x=x_1-\xi\sqrt{\Delta t},\quad y=x_1+\eta\sqrt{\Delta t}\]
gives
\begin{eqnarray}
&&J_{RL}(x_1,t,\Delta t)=\label{JRLexp}\\
&&\int_0^{\infty }d\xi \int_\xi^\infty\frac{d\zeta}{\sqrt{2\pi\Delta t}
\,\sigma(x_1,t)}\exp\left\{-\frac{\zeta^2}{2\sigma^2(x_1,t)}\right\} 
\,\Bigg{\{} p(x_{1},t)-\frac{\partial p(x_1,t)}{\partial t}
\Delta t \nonumber \\
&&+\sqrt{\Delta t}\Bigg{[} \left( \frac{\zeta ^{2}\left( \zeta -\xi
\right) \sigma ^{\prime }(x_{1},t)}{\sigma ^{3}(x_{1},t)}-\frac{\zeta
b(x_{1},t)}{\sigma ^{2}(x_{1},t)}-\frac{\sigma ^{\prime }(x_{1},t)\left(
\zeta -\xi \right) }{\sigma (x_{1},t)}\right)p(x_1,t)\nonumber\\
&&-\left(\zeta -\xi \right) p'(x_1,t)\Bigg{]}+O(\Delta
t)\Bigg{\}}.\nonumber
\end{eqnarray}
Since $p(x_{1},t)>0$, both $J_{LR}(x_{1},t)$ and $J_{RL}(x_{1},t)$ are
infinite, however, the net flux density is finite and is given by
\begin{eqnarray}
&&J_{\mbox{net}}(x_{1},t)=\lim_{\Delta t\rightarrow0}
\left\{J_{LR}(x_{1},t,\Delta t)-J_{RL}(x_{1},t,\Delta t)\right\} =\nonumber\\
&&-2\int_{0}^{\infty }d\xi \int_{\xi
}^{\infty }\frac{d\zeta }{\sqrt{2\pi \Delta t}\,\sigma (x_{1},t)}\exp
\left\{-\frac{\zeta^2}{2\sigma^2(x_{1},t)}\right\}\times\nonumber\\
&&\left[\left(\frac{\zeta^2\left(\zeta-\xi\right)\sigma'(x_{1},t)}
{\sigma^3(x_{1},t)}-\frac{\zeta b(x_{1},t)}{\sigma^2(x_{1},t)}-
\frac{\sigma'(x_{1},t)\left(\zeta -\xi \right)}{\sigma(x_{1},t)}\right)
p(x_{1},t)+\left( \zeta -\xi \right) p'(x_{1},t)\right]\nonumber\\
&&\quad\quad\quad\quad\quad=\left.-\frac{\partial}{\partial x}\frac{\sigma^2(x,t)}{2}
p(x,t)+b(x,t)p(x,t)\right|_{x=x_{1}}.  \label{JNET}
\end{eqnarray}
In deriving eq.(\ref{JNET}) use has been made of the following identities,
obtained by changing the order of integration,
\begin{eqnarray*}
\int_{0}^{\infty }d\xi \int_{\xi }^{\infty }\frac{\zeta ^{2}\left( \zeta
-\xi \right) \,d\zeta }{\sqrt{2\pi }\,\sigma }\exp \left\{ -\frac{\zeta ^{2}%
}{2\sigma ^{2}}\right\}  &=&\int_{0}^{\infty }\frac{\zeta ^{4}\,d\zeta }{2%
\sqrt{2\pi }\,\sigma }\exp \left\{ -\frac{\zeta ^{2}}{2\sigma ^{2}}\right\} =%
\frac{3\sigma ^{4}}{4} \\
\int_{0}^{\infty }d\xi \int_{\xi }^{\infty }\frac{\zeta \,d\zeta }{\sqrt{%
2\pi }\,\sigma }\exp \left\{ -\frac{\zeta ^{2}}{2\sigma ^{2}}\right\}
&=&\int_{0}^{\infty }\frac{\zeta ^{2}\,d\zeta }{\sqrt{2\pi }\,\sigma }\exp
\left\{ -\frac{\zeta ^{2}}{2\sigma ^{2}}\right\} =\frac{\sigma ^{2}}{2} \\
\int_{0}^{\infty }d\xi \int_{\xi }^{\infty }\frac{\left( \zeta -\xi \right)
\,d\zeta }{\sqrt{2\pi }\,\sigma }\exp \left\{ -\frac{\zeta ^{2}}{2\sigma ^{2}%
}\right\}  &=&\frac{1}{2}\int_{0}^{\infty }\frac{\zeta ^{2}\,d\zeta }{\sqrt{%
2\pi }\,\sigma }\exp \left\{ -\frac{\zeta ^{2}}{2\sigma ^{2}}\right\} =\frac{%
\sigma ^{2}}{4}.
\end{eqnarray*}
Equation (\ref{JNET}) is the classical expression for the probability (or
heat) current in diffusion theory \cite{Gardiner}. The Fokker-Planck
equation (\ref{FPEQ}) can be written in terms of the flux density function $%
J(x,t)$ in the conservation law form
\begin{equation}
\frac{\partial }{\partial t}p(x,t)=-\frac{\partial }{\partial x}J(x,t).
\label{divJ}
\end{equation}

Next, we calculate the uni-directional flux at the absorbing boundary $x=0$.
The absorbing boundary condition (\ref{abc}) implies that the pdf vanishes
for all $x\geq 0$ so that its right derivatives at the origin vanish. It
follows from eq.(\ref{JRLDt}) that 
\[
J_{RL}(0,t)=0. 
\]
On the other hand, eqs.(\ref{JLRDt}) and (\ref{JNET}) give 
\[
J(0,t)=J_{LR}(0,t)=\left. -\frac{\partial }{\partial x}\frac{\sigma ^{2}(x,t)%
}{2}p(x,t)\right| _{x=0}. 
\]
Since $\sigma ^{2}(x,t)>0$ and $p(x,t)>0$ for $x<0$, but $p(0,t)=0$, it
Since $\sigma ^2(x,t)>0$ and $p(x,t)>0$ for $x<0$, but $p(0,t)=0$, it
follows that $J(0,t)>0$. This means that there is positive flux into the
absorbing boundary so that the probability of trajectories that survive in
the region to the left of the absorbing boundary, $\int_{-\infty
}^0p(x,t)\,dx$, must be a decreasing function of time. This can be seen
directly from eq.(\ref{divJ}) by integrating it with respect to $x$ over the
ray $(-\infty ,0)$ and using the fact that $\lim_{x\rightarrow -\infty
}J(x,t)=0,$%
\[
\frac d{dt}\int_{-\infty }^0p(x,t)\,dx=-J(0,t)<0. 
\]

\vspace{3mm} \noindent
{\large {\bf 3. Current and uni-directional current in Feynman integrals}}\\
\renewcommand{\theequation}{3.\arabic{equation}} \setcounter{equation}{0}

To keep the calculations simple, we consider a particle with an infinite
potential for $x>0$ and its free propagation after the infinite potential is
turned off at time $t=0$. There is no analog to uni-directional current in
Feynman integrals due to the non-additivity of probability on sets of
trajectories. First, we examine the notion of current in the usual Feynman
integral. The current at a point is the net rate of change of probability on
one side of the point. That is, the current at $x=0$, say, is 
\begin{equation}
J(0,t)=\lim_{\Delta t\rightarrow 0}\frac 1{\Delta t}\int_{-\infty }^0\left[
\left| \psi (x,t+\Delta t)\right| ^{2\,}-\left| \psi (x,t)\right|
^{2\,}\right] \,dx.  \label{Ji0}
\end{equation}
According to the Feynman formalism, we write 
\[
\psi (x,t+\Delta t)=\sqrt{\frac m{2\pi i\hbar \Delta t}}\int_{-\infty
}^\infty \psi (y,t)\exp \left\{ \frac{im}{2\hbar \Delta t}(x-y)^2\right\}
\,dy 
\]
so that 
\begin{eqnarray*}
&&\int_{-\infty }^0\left| \psi (x,t+\Delta t)\right| ^{2\,}\,dx= \\
&&\int_{-\infty }^0\left| \sqrt{\frac m{2\pi i\hbar \Delta t}}\int_{-\infty
}^\infty \psi (y,t)\exp \left\{ \frac{im}{2\hbar \Delta t}(x-y)^2\right\}
\,dy\right| ^{2\,}\,dx.
\end{eqnarray*}
Thus the current is the difference between the population of trajectories
that propagates in the time interval $(t,t+\Delta t)$ from the entire line
into the interval $(-\infty ,0)$ and the population there at time $t$.

Expanding 
\begin{equation}
\psi (y,t)=\psi (x,t)+(y-x)\psi _x(x,t)+\frac 12(y-x)^2\psi _{xx}(x,t)+\cdot
\cdot \cdot ,  \label{parabola}
\end{equation}
we obtain 
\begin{eqnarray}
\int_{-\infty }^0 &&\left| \int_{-\infty }^\infty \sqrt{\frac m{2\pi i\hbar
\Delta t}}\psi (y,t)\exp \left\{ \frac{im}{2\hbar \Delta t}(x-y)^2\right\}
\,dy\right| ^{2\,}\,dx=  \label{dint1} \\
\int_{-\infty }^0 &&\left( \left| \psi (x,t)\right| ^2\,+\Re \mbox{e}\psi
(x,t)\bar{\psi}_{xx}(x,t)\frac{i\hbar \Delta t}m\right) \,dx+o(\Delta t).
\label{dint2}
\end{eqnarray}
It follows from integration by parts that 
\begin{equation}
J(0)=\frac\hbar m\Im\mbox{m}\int_{-\infty }^0\psi(x,t)\bar{\psi}_{xx}(x,t)\,dx=
\frac \hbar m\Im \mbox{m\thinspace }\psi (0,t)\bar{\psi}_x(0,t),  \label{j0}
\end{equation}
because $\psi _x(0,t)\bar{\psi}_x(0,t)$ is real valued. Equation (\ref{j0})
is identical to the Schr\"{o}dinger current. This derivation seems to be new.

Next, we consider the case that the wave function at time $t$ vanishes
outside the interval $[-a,0]$. As mentioned above, this situation arises,
for example, if up to time $t$ there is an infinite potential for $x>0$ and $%
x<-a$ and the potential for $x>0$ is turned off instantaneously at time $t$.
This is a mathematical idealization of various physical situations (see,
e.g., Chapter 6). Since $\psi (0,t)=0$, the Schr\"{o}dinger current at the
point $x=0$ vanishes at time $t$, according to eq.(\ref{j0}). Yet, there is
probability flux across $x=0$. To see this, we evaluate the rate of
population change, (\ref{Ji0}). In the case at hand the second term in the
integrand of (\ref{Ji0}) vanishes at time $t$. We assume first that $x=-a$
is a reflecting wall, that is, $\psi (x,t)=0$ for $x\leq -a$ for some
positive $a$. Thus there is no propagation across $x=-a$. The case $a=\infty 
$ is considered in Section 5 below.

The probability of propagating from a given interval $[-a,0]$ into the ray $%
[0,\infty )$ in the time interval $(t,t+\Delta t)$, starting with the wave
function $\psi (x,t)$ in the interval $[-a,0]$ and $0$ outside, is given by 
\begin{equation}
P_{\mbox{out}}=\int_0^\infty \left| \psi (y,t+\Delta t)\right| ^2dy.
\label{Pout}
\end{equation}
The wave function at time $t+\Delta t$ is given by the free propagator 
\begin{equation}
\psi (y,t+\Delta t)=\sqrt{\frac m{2\pi i\hbar \Delta t}}\int_{-a}^0\psi
(x,t)\exp \left\{ \frac{im(x-y)^2}{2\hbar \Delta t}\right\} dx.
\label{psiytdt}
\end{equation}
The uni-directional current from $[-a,0]$ into the ray $[0,\infty )$ is the
current 
\begin{eqnarray}
J_{LR}(0,t)=\lim_{\Delta t\rightarrow 0}\frac 1{\Delta t}\int_0^\infty
\left| \psi (x,t+\Delta t)\right| ^{2\,}\,dx.  \label{JLR0}
\end{eqnarray}
Note that if trajectories are not truncated for $x>0$, the wave function at
time $t+\Delta t$ no longer vanishes at $x=0$ and so does the
Schr\"{o}dinger current.\newpage

\noindent
{\large {\bf 4. Short time propagation}}\newline
\renewcommand{\theequation}{4.\arabic{equation}} \setcounter{equation}{0}

To estimate the uni-directional current $J_{LR}(0,t)$ for short times, we
first estimate the integral in eq.(\ref{JLR0}) for small $\Delta t$. We
begin with an initial wave function $\psi (x,0)$ that is a polynomial 
\[
Q(x)=\sum_{j=1}^Nq_jx^j 
\]
in the interval $[-a,0]$, such that $Q(-a)=Q(0)=0$ and $\psi (x,0)=0$
otherwise, the free propagation from the interval $[-a,0]$ is given by 
\[
\psi (y,\Delta t)=\sqrt{\frac m{2\pi i\hbar \Delta t}}\int_{-a}^0Q(x)\exp
\left\{ \frac{im(x-y)^2}{2\hbar \Delta t}\right\} \,dx. 
\]
The boundary condition at the left end of the support of $\psi (x,0)$ is
written explicitly as 
\[
\sum_{j=1}^Nq_j\left( -a\right) ^j=0. 
\]
Setting 
\[
\alpha =\frac{\hbar \Delta t}m, 
\]
the probability mass propagated freely into the positive axis in time $%
\Delta t$ is given by 
\[
\int_0^\infty \left| \psi \left( y,t+\Delta t\right) \right| ^2dy=\frac
1{2\pi \alpha }\int_0^\infty \left| \int_{-a}^0Q(x)e^{i\left( x-y\right)
^2/2\alpha }\,dx\right| ^2dy. 
\]
We change variables by setting $x=\sqrt{\alpha }\xi ,\quad y=\sqrt{\alpha }%
\eta ,\quad \xi =\zeta +\eta $ to get 
\begin{equation}
\int_0^\infty \left| \psi \left( y,t+\Delta t\right) \right| ^2dy=\frac{%
\alpha ^{1/2}}{2\pi }\int_0^\infty \left| \int_{-a/\sqrt{\alpha }-\eta
}^{-\eta }Q\left( \sqrt{\alpha }\left( \zeta +\eta \right) \right) e^{i\zeta
^2/2}\,d\zeta \right| ^2d\eta .  \label{integral}
\end{equation}
First, we evaluate the inner integral, 
\[
I_N\left( \eta \right) =\sum_{j=1}^Nq_j\sqrt{\alpha ^j}\int_{-a/\sqrt{\alpha 
}-\eta }^{-\eta }\left( \zeta +\eta \right) ^je^{i\zeta ^2/2}\,d\zeta . 
\]
Integration by parts gives 
\begin{eqnarray}
I_N\left( \eta \right) &=&-i\sum_{j=1}^Nq_j\sqrt{\alpha ^j}\int_{-a/\sqrt{%
\alpha }-\eta }^{-\eta }\left( \zeta +\eta \right) ^j\frac{de^{i\zeta ^2/2}}%
\zeta  \label{I_N} \\
&=&-i\sum_{j=1}^Nq_j\sqrt{\alpha ^j}\left[ \left( \frac{-a}{\sqrt{\alpha }}%
\right) ^j\frac{e^{i\left( a/\sqrt{\alpha }+\eta \right) ^2/2}}{a/\sqrt{%
\alpha }+\eta }\right] +i\sum_{j=1}^Nq_j\sqrt{\alpha ^j}\int_{-a/\sqrt{%
\alpha }-\eta }^{-\eta }e^{i\zeta ^2/2}\frac d{d\zeta }\frac{\left( \zeta
+\eta \right) ^j}\zeta d\zeta  \nonumber \\
&=&i\sum_{j=1}^Nq_j\sqrt{\alpha ^j}\int_{-a/\sqrt{\alpha }-\eta }^{-\eta
}e^{i\zeta ^2/2}\left( \frac{j\left( \zeta +\eta \right) ^{j-1}}\zeta -\frac{%
\left( \zeta +\eta \right) ^j}{\zeta ^2}\right) d\zeta ,  \nonumber
\end{eqnarray}
because 
\[
\sum_{j=1}^Nq_j\sqrt{\alpha ^j}\left( \frac{-a}{\sqrt{\alpha }}\right)
^j=\sum_{j=1}^Nq_j\left( -a\right) ^j=0. 
\]
For $j=1$, we obtain 
\[
-iq_1\sqrt{\alpha }\eta \int_{-a/\sqrt{\alpha }-\eta }^{-\eta }\frac{%
e^{i\zeta ^2/2}}{\zeta ^2}d\zeta . 
\]
The function 
\[
\eta \int_{-a/\sqrt{\alpha }-\eta }^{-\eta }\frac{e^{i\zeta ^2/2}}{\zeta ^2}%
d\zeta 
\]
is square integrable and its integral is independent of $\alpha $ to leading
order. Indeed, the integral 
\[
\int_{-\infty }^{-\eta }\frac{e^{i\zeta ^2/2}}{\zeta ^2}d\zeta 
\]
exists and near $\eta =0\ $it is bounded by 
\[
\left| \int_{-\infty }^{-\eta }\frac{e^{i\zeta ^2/2}}{\zeta ^2}d\zeta
\right| \leq \int_{-\infty }^{-\eta }\frac{d\zeta }{\zeta ^2}=\frac 1\eta 
\]
so that 
\[
\left| \eta \int_{-a/\sqrt{\alpha }-\eta }^{-\eta }\frac{e^{i\zeta ^2/2}}{%
\zeta ^2}d\zeta \right| \leq \eta \int_{-\infty }^{-\eta }\frac{d\zeta }{%
\zeta ^2}=1. 
\]
For large $\eta $, we have the asymptotic limit 
\[
\int_{-\infty }^{-\eta }\frac{e^{i\zeta ^2/2}}{\zeta ^2}d\zeta \sim i\frac{%
e^{i\eta ^2/2}}{\eta ^3}, 
\]
as is easily seen from l'Hospitale's rule, so that 
\[
\left| \eta \int_{-a/\sqrt{\alpha }-\eta }^{-\eta }\frac{e^{i\zeta ^2/2}}{%
\zeta ^2}d\zeta \right| ^2\leq \frac 1{\eta ^4}. 
\]
This means that the function $\left| \eta \int_{-a/\sqrt{\alpha }-\eta
}^{-\eta }\frac{e^{i\zeta ^2/2}}{\zeta ^2}d\zeta \right| ^2$ is integrable.
It follows that its contribution to the integral (\ref{integral}) is to
leading order 
\begin{eqnarray}
\frac{\alpha ^{3/2}}{2\pi }\left| q_1\right| ^2\int_0^\infty \left| \eta
\int_{-\infty }^{-\eta }\frac{e^{i\zeta ^2/2}}{\zeta ^2}d\zeta \right|
^2d\eta =O\left( \alpha ^{3/2}\right) . \label{q1a}
\end{eqnarray}
Now, we consider the term $j=2$ : 
\[
iq_2\alpha \int_{-a/\sqrt{\alpha }-\eta }^{-\eta }e^{i\zeta ^2/2}\left( 1-%
\frac{\eta ^2}{\zeta ^2}\right) d\zeta . 
\]
Proceeding as above, we find that both terms in the integral are uniformly
square integrable functions of $\eta $ for all $\alpha >0$ sufficiently
small. It follows that the contribution of the term $j=2$ to the integral (%
\ref{integral}) is $O\left( \alpha ^{5/2}\right) $. The mixed term involving 
$q_1q_2$ contributes $O\left( \alpha ^2\right) $.

Next, we consider the third order term: 
\begin{eqnarray}
&&\int_{-a/\sqrt{\alpha }-\eta }^{-\eta }\left( \zeta +\eta \right) ^2\frac{%
2\zeta -\eta }{\zeta ^3}de^{i\zeta ^2/2}=  \nonumber \\
&&-\left( -\frac{a}{\sqrt{\alpha}}\right) ^2\frac{2\left( a/\sqrt{\alpha}%
+\eta \right) -\eta }{\left( -a/\sqrt{\alpha }-\eta \right) ^3}e^{i\left( a/%
\sqrt{\alpha }+\eta \right) ^2/2} -\int_{-a/\sqrt{\alpha }-\eta }^{-\eta
}e^{i\zeta ^2/2}\frac d{d\zeta }\left[ \left( \zeta +\eta \right) ^2\frac{%
2\zeta -\eta }{\zeta ^3}\right] d\zeta  \nonumber \\
&&=O\left( \frac 1\alpha \right) .  \label{q3}
\end{eqnarray}
This term has a pre-factor of $q_3\alpha ^{3/2}$ so that its contribution to
(\ref{integral}) is $O\left( \alpha ^{3/2}\right) $. Proceeding by
induction, we find that all terms contribute $O\left( \alpha ^{3/2}\right) $
to (\ref{integral}). It follows that 
\begin{eqnarray}
\int_0^\infty \left| \psi \left( y,t+\Delta t\right) \right| ^2dy=O\left(
\left( \frac{\hbar \Delta t}m\right) ^{3/2}\right).  \label{mass}
\end{eqnarray}

If $a=\infty$ and all Fresnel-type integrals are interpreted as the limits 
\begin{eqnarray*}
\int_{-\infty}^yf(x)e^{ix^2/2}\,dx=\lim_{\epsilon\downarrow0}
\int_{-\infty}^yf(x)e^{(-\epsilon+i)x^2/2}\,dx,
\end{eqnarray*}
the term $O\left(1/\alpha\right)$ in eq.(\ref{q3}) is replaced by $O(1)$. It
follows that the higher order terms of the polynomial contribute higher
order terms in the expansion of $I_N$ in powers of $\alpha$.

Obviously, if the polynomial is replaced by an analytic function that
vanishes at the ends of the interval, the result remains unchanged.
Furthermore, if $Q(x)$ is a square integrable analytic function on the
negative axis, eq.(\ref{mass}) holds.

The asymptotic estimate (\ref{mass}) is valid when higher order terms can be
neglected relative to lower order terms. To get an explicit bound on $\Delta
t$ from this condition, we write $\psi (x,t)$ as a series of eigenfunctions 
\begin{eqnarray}
\psi (x,t)=\sum_{j=1}^\infty \psi _j(x)\exp \left\{ -\frac{iE_nt}\hbar
\right\},\label{psixt=}
\end{eqnarray}
where $\psi _n(x)$ are eigenfunctions that satisfy the boundary condition $%
\psi _n(0)=0$ and $E_n$ are the corresponding eigen energies, to find that 
\[
\frac{\partial ^j\psi (0,t)}{\partial x^j}=j!q_j(t)
\]
and 
\begin{eqnarray}
\left| \frac{\partial ^{2j+1}\psi _n(0)}{\partial x^{2j+1}}\right| 
&=&\left| \frac{-2mE_n}{\hbar ^2}\right| ^j\left| \frac{\partial \psi _n(0)}{%
\partial x}\right|   \label{psipen} \\
\left| \frac{\partial ^{2j}\psi _n(0)}{\partial x^{2j}}\right|  &=&\left| 
\frac{-2mE_n}{\hbar ^2}\right| ^j\left| \psi _n(0)\right| =0.  \nonumber
\end{eqnarray}

The asymptotic evaluation of the integrals for $y$ near or at $0$ gives that
the coefficient of $q_j$ in the expansion is $O\left( \left( \frac{\hbar
\Delta t}m\right) ^{j+1/2}\right) $. It follows that the condition for the
validity of the expansion is that 
\begin{eqnarray}
\frac{\hbar \Delta t}m\ll \left| \frac{q_{2j+1}(t)}{q_{2j+3}(t)}\right| 
\label{qq}
\end{eqnarray}
for all $j\geq 0$. Using eqs.(\ref{psixt=})-(\ref{psipen}) in (\ref{qq}), we
obtain that the condition for the validity of the expansion is 
\begin{equation}
\frac{\hbar \Delta t}m\ll \left| \frac{(2j+1)(2j+3)\sum_{n=1}^\infty \left( 
\frac{-2mE_n}{\hbar ^2}\right) ^j\frac{\partial \psi _n(0)}{\partial x}%
e^{-iE_nt/\hbar }}{\sum_{n=1}^\infty \left( \frac{-2mE_n}{\hbar ^2}\right)
^{j+1}\frac{\partial \psi _n(0)}{\partial x}e^{-iE_nt/\hbar }}\right|
\label{condition}
\end{equation}
for all $j\geq 1$.

If, for example, the initial wave function is a single eigenfunction, the
condition (\ref{condition}) reduces to 
\begin{eqnarray}
E_n\Delta t\ll \hbar . \label{uncert}
\end{eqnarray}
The analysis of the continuous spectrum case is identical. The summation
with respect to $n$ in the condition (\ref{condition}) is replaced by
integration with respect to $n$.

It follows from eq.(\ref{mass}) that for short times 
\begin{eqnarray*}
J_{LR}(0,t)=O(\sqrt{t})
\end{eqnarray*}
so that the population in $y>0$ increases as $O(t^{3/2})$ for short times.
Obviously, once $\psi (0,t)\neq 0$, the current becomes the usual
Schr\"{o}dinger current.

The probability mass that propagates in time $\Delta t$ beyond a fixed point 
$c>0$ can be found from the above expansions. This probability is defined as 
\begin{eqnarray}
P_c=\int_c^\infty \left| \psi (y,t+\Delta t)\right|^2dy= \frac{\alpha^{1/2}}{%
2\pi}\int_{c/\sqrt{\alpha}}^\infty \left|\int_{-a/\sqrt{\alpha }-\eta
}^{-\eta }Q\left( \sqrt{\alpha }\left( \zeta +\eta \right) \right) e^{i\zeta
^2/2}\,d\zeta \right| ^2d\eta,  \label{Pc}
\end{eqnarray}
rather than (\ref{integral}). The individual terms in the expansion (\ref
{I_N}) are estimated as above with the obvious changes. Inequality (\ref{q1a}%
) becomes 
\begin{eqnarray}
\frac{\alpha^{3/2}}{2\pi}\left|q_1\right|^2\int_{c/\sqrt{\alpha}%
}^\infty\left|\eta \int_{-\infty }^{-\eta }\frac{e^{i\zeta^2/2}}{\zeta^2}%
d\zeta \right| ^2d\eta =O\left( \alpha ^3\right) .  \label{q1a3}
\end{eqnarray}
The same estimate applies to all terms in the expansion. It follows that 
\begin{eqnarray}
P_c=O\left(\left(\frac{\alpha}{c}\right)^3\right).  \label{P_c}
\end{eqnarray}
This expansion is valid for $c\geq\sqrt{\alpha}$. For example, if $%
c=O\left(\alpha^{1/3}\right)$, we obtain $P_c=O\left(\alpha^2\right)$. For $%
c=O\left(\sqrt{\alpha}\right)$ the result (\ref{mass}) is recovered.

Now, we consider an initially discontinuous wave function 
\[ \psi \left( x,0\right) =\left\{ \begin{array}{lll}
Q(x) &  & \mbox{for }x<0 \\ &  &  \\ 0 &  & \mbox{for }x\geq 0
\end{array} \right. , \]
where $Q(x)=\sum_{j=0}^Nq_jx^j$ vanishes at $x=-a$ and $Q\left( 0\right)
\neq 0$. First, we consider the propagation of the term $q_0$. Its contribution to
the propagated wave function is 
\[
\frac{\alpha ^{1/2}}{2\pi }\int_0^\infty \left| q_0\int_{-a/\sqrt{\alpha }%
-\eta }^{-\eta }e^{i\zeta ^2/2}\,d\zeta \right| ^2\,d\eta =O\left( \alpha
^{1/2}\right) .
\]
This gives rise to an infinite uni-directional current at the point of
discontinuity. Next, we calculate the probability propagated beyond a
distance $c>0$ from the discontinuity. Again, the term $q_0$ contributes
\begin{eqnarray}
P_c=\frac{\alpha ^{1/2}}{2\pi }\int_{c/\sqrt{\alpha }}^\infty \left| q_0\int_{-a/%
\sqrt{\alpha }-\eta }^{-\eta }e^{i\zeta ^2/2}\,d\zeta \right| ^2\,d\eta =%
\frac{\left| q_0\right| ^2\alpha }{2\pi c}+O\left( \frac{\alpha ^{3/2}}%
c\right) .\label{Pc1}
\end{eqnarray}
It follows that for $c=O\left( \alpha ^{1/2-\varepsilon }\right) $ the
propagated probability is $O\left( \alpha ^{1/2+\varepsilon }\right) $, so
that the resulting uni-directional current is infinite for $0<\varepsilon
<1/2$. If $c=O\left( 1\right) $, the propagated probability is $O\left(
\Delta t\right) $ and gives rise to a finite current.

The analysis of the propagation of a polynomial shows that the other terms
in the polynomial contribute higher order terms to the propagated
probability.\\

\noindent
{\large {\bf 5. Discussion and summary}}\newline
\renewcommand{\theequation}{5.\arabic{equation}} \setcounter{equation}{0}

The expression (\ref{mass}) can be applied to the following experiment. A
particle with energy $E_n$ is released between two perfectly reflecting
walls placed at $x=0$ and $x=-a$ ($a>0$). A perfectly absorbing detector is
placed at $x_0>0$. The reflecting wall at $x=0$ is removed instantaneously
for a time interval of length $\Delta t$ and is then instantaneously
reinstated. If the particle crosses $0$ in this time interval, it gets
registered by the detector. According to eq.(\ref{mass}), if this experiment
is repeated $N$ times, the number of particles registered by the detector
will be proportional to $N\left( \Delta t\right) ^{3/2}$, if $\Delta t$
satisfies the condition (\ref{uncert}).

Note that the condition (\ref{uncert}) in this context has no relation to
the energy-time uncertainty principle because $\Delta t$ is not related to a
measurement of the particle's energy. The removal and reinstatement of the
reflecting wall changes the energy of the particle so that it is not known
what it is after the wall is reinstated.

A possible physical approximate realization of this experiment consists in
turning off and on again a detecting device, e.g., by illuminating
the detection region (and using e.g., the Compton effect), a time 
interval of length $\Delta t$. According 
to eq.(\ref{mass}), the survival probability on the left of the barrier is 
\begin{eqnarray*}
S\left(\Delta t\right)=1-c\left(\Delta t\right)^{3/2},
\end{eqnarray*}
where $c$ is a constant. If this experiment is run on $N$ identical systems
with $N\Delta t=T$ and $\Delta t$ satisfies the condition (\ref{uncert}),
the probability that all survive by time $\Delta t$, denoted $S_T(N)$, is
\begin{eqnarray*}
S_T(N)&=&\left(1-c\left(\Delta t\right)^{3/2}\right)^N\approx \exp\left\{-%
\frac{c}{\sqrt{N}}\right\}\\
&=&1-\frac{c}{\sqrt{N}},\quad N>>1.
\end{eqnarray*}
The expected number of systems that decay is 
\begin{eqnarray*}
\langle N\rangle=O\left(N\Delta t^{3/2}\right)=O\left(T\sqrt{\Delta
t}\right)\to0\quad\mbox{for $N\gg1$}.
\end{eqnarray*}

The result (\ref{mass}) is similar to that obtained in time dependent
perturbation theory, known as the Zeno effect \cite{Peres,Aharonov},
where the probability that an irreversible decay will occur before a
short time $t$ is $O\left(t^2\right)$. As in the Zeno effect, 
the law (\ref{mass}) indicates that propagation into a detection
region under continuous observation makes it impossible
for a particle to be observed. This phenomenon is referred to as 
the freezing of a particle in its initial state. This apparent 
paradox disappears in quantum theory with a measuring device, as 
shown in \cite{PHD,measurement}.

We return now to the ideal detection experiment by illuminating the
detection region $x>x_0$ for a short time $\Delta t$ at time intervals
$\Delta t$ apart. The detection probability is $P_c$, given in
eq.(\ref{Pc}). If $x_0=0$, the detection probability is
$O\left(\alpha^{3/2}\right)$ per illumination pulse. The result
remains the same if the illuminated region is the interval $(0,c)$,
where $c=O\left(\sqrt{\alpha}\right)$. If, however, one
illumination pulse covers the region $x>0$ and the following one
covers $x>c$, where $c=O(1)$ for $\alpha$ satisfying eq.(\ref{qq}) or,
equivalently, (\ref{uncert}), the conditional probability of 
detecting the particle in the second pulse, given that it was not 
detected in the first one, is $O\left(\alpha^3\right)$. The former
result means that the width of the illuminated region has to be at
least $O\left(\sqrt{\alpha}\right)$ to achieve the maximal order of
magnitude of the probability of detection per pulse.

The result eq.(\ref{Pc1}) can be applied to the following ideal
measurement experiment. A discontinuous wave function is created by
introducing a potential $V(x)=\delta'(x)$ \cite{Cheon}. If the
illumination region is $x>0$, the probability of detection per pulse
is $O\left(\sqrt{\alpha}\right)$. If the the regions $x>0$ and $x>c$
are illuminated alternatively, the conditional probability of detecting
the particle in the second pulse, given that it was not observed
during the first one, is $O\left(\alpha\right)$, according to
eq.(\ref{Pc1}). This means, that if in
the second pulse the illumination of the region $x>c$ is kept
forever, the conditional survival probability  $S(t)$ (the probability of
not observing the particle by time $t$ after the beginning of the
second pulse)  is
\begin{eqnarray}
S(t)=O\left(\exp\left\{-\gamma \int_0^t\left|q_0(t')\right|^2\,dt'\right\}
\right)\label{S(t)}
\end{eqnarray}
for some $\gamma>0$. The probability propagated across the boundary of the support in
time $\Delta t$ is $O\left( \Delta t^{1/2}\right) $ and the
initial uni-directional current is $O\left(\Delta t^{-1/2}\right)$. 
This is an anti-Zeno effect \cite{Kaulakys}. It means that in $N$ identical
systems observed for time $\Delta t=T/N$ the the probability that all survive
by time $\Delta t$ is
\begin{eqnarray*}
S_T(N)&=&\left(1-c\left(\Delta t\right)^{1/2}\right)^N\approx \exp\left\{-%
{c}{\sqrt{N}}\right\}\to0\quad\mbox{for $N\gg1$}.
\end{eqnarray*}
The expected number of systems that decay is
\begin{eqnarray*}
\langle N\rangle=O\left(N\Delta t^{1/2}\right)=O\left(\sqrt{TN}\right)\to\infty
\quad\mbox{for $N\gg1$}.
\end{eqnarray*}

It should be remarked that the leading order short time asymptotics 
is unaffected by the presence of a finite potential beyond the boundary
of the support of the initial wave function. This suggests the
possibility that the detection region beyond the support of the initial
wave function can be characterized by a potential without essentially
changing the above result.

In summary, we compared the notions of net and uni-directional fluxes in the
Wiener and Feynman integrals. At points where the density does not vanish
the uni-directional fluxes are infinite, though the net flux is finite and is
given by the traditional expressions for flux in the diffusion and
Schr\"{o}dinger equations. At points where the density vanishes, for example
at certain types of boundaries (absorbing for Wiener trajectories and
at the boundary of the support of the wave function) 
the uni-directional fluxes are finite.
In the Wiener integral the uni-directional flux at an absorbing boundary does
not vanish, resulting in a decay of the total population at an exponential
rate. In contrast, if a reflecting boundary for the Feynman integral is
removed at time $t=0$, the flux across the boundary increases as $O(\sqrt{t}%
) $. If Feynman trajectories that propagate into a boundary are
instantaneously absorbed there, the flux at such a boundary is proportional
to the square of the local gradient of the wave function 
\cite{absorption}.\\

\noindent
{\bf Acknowledgment:} The authors wish to thank Y. Aharonov and B.
Reznik for useful discussions.

\end{document}